\documentclass[prl,aps,twocolumn]{revtex4}
\usepackage{graphicx}
\usepackage{subfigure}
\usepackage{bm}
\usepackage{amsmath}
\begin{document} 
\title{Resonance-assisted tunneling in 
three degrees of freedom without discrete symmetry}
\author{Srihari Keshavamurthy\footnote{Permanent address: Department of
Chemistry, Indian Institute of Technology, Kanpur, U.P. 208016, India.}}
\affiliation{School of Mathematics, University of Bristol, University Walk,
Bristol, BS8 1TW, United Kingdom}
\date{\today}
\begin{abstract}
We study dynamical tunneling
in a near-integrable Hamiltonian
with three degrees of freedom. The model Hamiltonian does not
have any discrete symmetry. Despite this lack of symmetry we
show that the
mixing of near-degenerate quantum states 
is due to dynamical tunneling 
mediated by the nonlinear resonances in the classical
phase space. Identifying the key resonances allows us to 
suppress the dynamical tunneling via the addition of weak
counter-resonant terms.
\end{abstract}
\maketitle

Tunneling is a phenomenon that is forbidden
by classical mechanics but allowed by quantum mechanics. 
In general, any
flow of quantum probability between 
(approximately) equivalent yet classically disconnected regions
constitutes tunneling. The classical regions could be disconnected
due to barriers in coordinate space, momentum space or, more
generally, in the classical phase space. In the cases where tunneling
occurs despite the absence of obvious energetic barriers it is called
{\em dynamical tunneling}\cite{davhel}; the barriers now 
arise due to certain exact or approximate constants of
the motion and hence are naturally identified in the underlying classical
phase space. 
Considerable 
theoretical\cite{davhel,cats,cat1,frido,rat1,rat2,ozo,ejhsar,elt,self} 
and experimental\cite{expt} works have established that
tunneling between quantum states localized on two symmetry-related 
regions in the phase space
can be strongly influenced by the classical
stochasticity (chaos-assisted tunneling\cite{cats}) and/or by the intervening
nonlinear resonances (resonance-assisted tunneling\cite{rat2}). 
In the former case, phase space is mixed regular-chaotic and
the splittings show marked dependence on the nature of the chaotic
states which couple to the tunneling doublets\cite{cats,cat1,frido}. 
In the latter case
with near-integrable phase space, the splittings
depend delicately on the various resonance islands bridging
the degenerate states\cite{rat1,rat2,ozo,ejhsar,elt,self}. 
Clearly, a quantitative semiclassical theory, still elusive,
requires one to identify 
key structures in the phase space on which the 
theory is to be based. In this regard there is increasing 
evidence\cite{elt,self}
that the classical nonlinear resonances might 
play a central role
in near-integrable as well as mixed phase space situations.

However, most of the studies thus far have been on two degrees of freedom (dof)
systems with discrete symmetries\cite{excep}. 
Does the resonance-assisted tunneling
viewpoint hold in systems with three or more dof
which lack discrete symmetries?
The main motivation for our study comes from 
suggestions\cite{ejhsar} put forward in the molecular
context - can dynamical tunneling provide a route for mixing
between near-degenerate states and hence energy flow between
regions supporting qualitatively different types of motion?
In addition, notwithstanding
the difficulties associated with visualizing the multidimensional
phase space, dynamics
in three or more dof has features that cannot manifest
in the systems studied up until now\cite{licht}.
In this letter we attempt to
understand dynamical tunneling in a 
model nonsymmetric, near-integrable three dof
system. We show that mixing of near-degenerate states
occurs via dynamical tunneling 
mediated by nonlinear resonances and the mixing can be suppressed 
by adding weak counter-resonant terms.

We study the Hamiltonian 
\begin{equation}
H=H_{0} 
+ \sum_{r} K_{{\bf m}_{r}}[(a_{1}^{\dagger})^{\alpha_{r}}
(a_{2})^{\beta_{r}} (a_{3})^{\gamma_{r}} (a_{4})^{\delta_{r}}
+ {\rm h.c.}], \label{qham} 
\end{equation}
describing four coupled modes $j=1,2,3,4$ with 
\begin{equation}
H_{0}=\sum_{j}(\omega_{j}n_{j}+x_{jj}n_{j}^{2})
+ \sum_{j<k} x_{jk} n_{j}n_{k},
\end{equation}
and $H_{0}|{\bf n}\rangle = E_{\bf n}^{0}|{\bf n}\rangle$.
Although eq.~\ref{qham} has been inspired in the molecular
context\cite{quack}, similar multiresonant Hamiltonians arise in a 
variety of systems\cite{simil1}. 
The occupation number of the $j^{th}$ mode, 
$n_{j}\equiv a_{j}^{\dagger}a_{j}$, is expressed in terms of the
harmonic creation ($a_{j}^{\dagger}$) and destruction ($a_{j}$)
operators.
The perturbations are characterized by 
${\bf m}_{r} = (\alpha_{r},-\beta_{r},-\gamma_{r},-\delta_{r})$ with
strengths $K_{{\bf m}_{r}}$. 
The classical limit of eq.~\ref{qham}, generated via the
correspondence $a_{j} \leftrightarrow \sqrt{I_{j}}\exp(i\theta_{j})$,
is the following Hamiltonian:
\begin{equation}
{\cal H}({\bf I},{\bm \theta})={\cal H}_{0}({\bf I})
+2 \epsilon \sum_{r} K_{{\bf m}_{r}}
\sqrt{I_{1}^{\alpha_{r}} I_{2}^{\beta_{r}} I_{3}^{\gamma_{r}}
I_{4}^{\delta_{r}}} \cos({\bf m}_{r} \cdot {\bm \theta}).
\label{cham}
\end{equation}
$({\bf I},{\bm \theta})$ are the classical action-angle variables
of ${\cal H}_{0}$ and hence the perturbations 
correspond to classical nonlinear resonances. 
The parameter $\epsilon$ has been introduced for a perturbative
analysis (see below).
We restrict ourselves to 
three perturbations
${\bf m}_{1}=(1,-2,0,0)$, ${\bf m}_{2}=(1,-1,-1,0)$, and
${\bf m}_{3}=(1,-1,0,-1)$. 
This allows for a clear study of the role of the specific resonances
in dynamical tunneling. 
The existence of a conserved quantity 
$P=n_{1}+(n_{2}+n_{3}+n_{4})/2$, with the classical
analog $P_{c}=I_{1}+(I_{2}+I_{3}+I_{4})/2$,
implies that the 4-mode system has effectively three
dof. 
The eigenstates, eigenvalues, and the resulting mean level spacing
of $H$ are denoted by $|\alpha \rangle$,
$E_{\alpha}$ and $\Delta E$ respectively.
In the units appropriate for the model Hamiltonian\cite{quack} the
Heisenberg time is given by $\tau_{H}=(2\pi c \Delta E)^{-1}$ with $c$ 
being the speed of light. 

We are interested in the fate of a set of near-degenerate
zeroth-order states in the presence of 
weak perturbations, $K_{{\bf m}_{r}}/\Delta E \equiv k_{{\bf m}_{r}} < 1$. 
Consider
states $|{\bf n}\rangle$, $|{\bf n}'\rangle$, $|{\bf n}"\rangle, \ldots$
such that $E_{\bf n}^{0} \approx E_{\bf n'}^{0} \approx E_{\bf n"}^{0}
\approx \ldots$ with average energy $\bar{E}$ and 
$E_{\bf n}^{0} \in (\bar{E}-\Delta E/2,\bar{E}+\Delta E/2)$. 
Certain states, 
among the set of near-degenerate states, 
mix since they are directly connected to each other via
one of the perturbations. The nonlinear resonances in eq.~\ref{cham}
do mediate the mixing via dynamical tunneling.
However, in this
work we will focus on states that are not directly
coupled by the resonances in eq.~\ref{cham} in order to show that
even very weak, induced resonances can lead to substantial mixing
that can be associated with dynamical tunneling.
Quantum mechanically, the extent of mixing of a zeroth-order state
$|{\bf n}\rangle$ can be gauged by computing the survival probability
$P_{{\bf n}Q}(t)$ and the inverse participation ratio (IPR) 
$\sigma_{\bf n}$,
\begin{eqnarray}
P_{{\bf n}Q}(t) &=& |\langle {\bf n}|e^{-iHt/\hbar}|{\bf n}\rangle|^{2}
= \sum_{\alpha,\beta} p_{{\bf n}\alpha} p_{{\bf n}\beta} 
e^{-i\omega_{\alpha \beta}t}, \\
\sigma_{\bf n} &=& \lim_{T \rightarrow \infty} \frac{1}{T} \int_{0}^{T}
P_{{\bf n}Q}(t) dt = \sum_{\alpha} p^{2}_{{\bf n}\alpha}, \label{ipr}
\end{eqnarray}
with $p_{{\bf n}\alpha}=|\langle \alpha|{\bf n}\rangle|^{2}$
and $\omega_{\alpha \beta}=(E_{\alpha}-E_{\beta})/\hbar$.
If $\sigma_{\bf n} \ll 1$ then $|{\bf n}\rangle$ mixes extensively
with other zeroth-order states.

\begin{figure}
\includegraphics*[height=3in,width=3in]{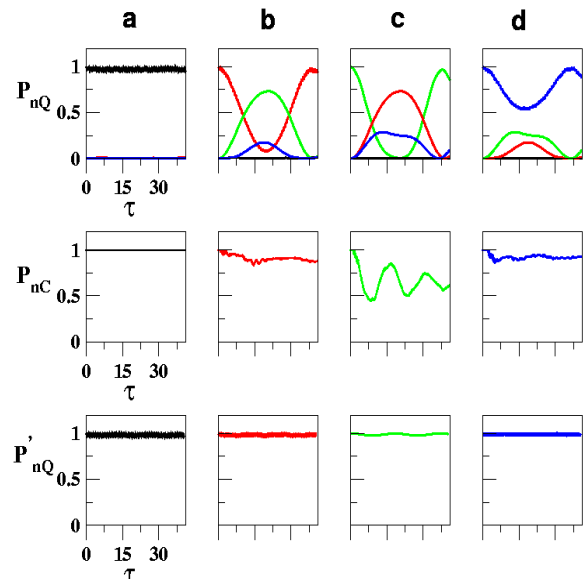}
\caption{Quantum survival probability (top row) $P_{{\bf n}Q}$
of the states
$|{\bf n}\rangle=|0,11,1,4\rangle \equiv |a\rangle$ (black),
$|0,11,2,3\rangle \equiv |b\rangle$ (red),
$|0,12,2,2\rangle \equiv |c\rangle$ (green),
and $|0,13,1,2\rangle \equiv |d\rangle$ (blue).
Time $\tau$ is measured in units of the Heisenberg time and
$0.5 K_{{\bf m}_{1}} = K_{{\bf m}_{2}} = K_{{\bf m}_{3}} \approx
0.2 \Delta E$. $\Delta E \approx 4.44$ cm$^{-1}$ for eq.~\ref{qham}
with $P=8$. The cross probabilities $|\langle {\bf n}'|{\bf n}(t)\rangle|^{2}$
are also shown.
The middle row shows the classical
analog, $P_{{\bf n}C}$, of $P_{{\bf n}Q}$.
$P_{{\bf n}C}$ indicates
trapped motion and is qualitatively different from $P_{{\bf n}Q}$.
The bottom row shows the quantum
$P^{'}_{{\bf n}Q}$ with the
addition of weak counter-resonant terms (eq.~\ref{mqham}).
The dynamical tunneling
essentially shuts down, proving the importance of the induced
resonances.}
\label{fig1}
\end{figure}

\begin{figure}
\includegraphics*[height=3.0in,width=3.0in]{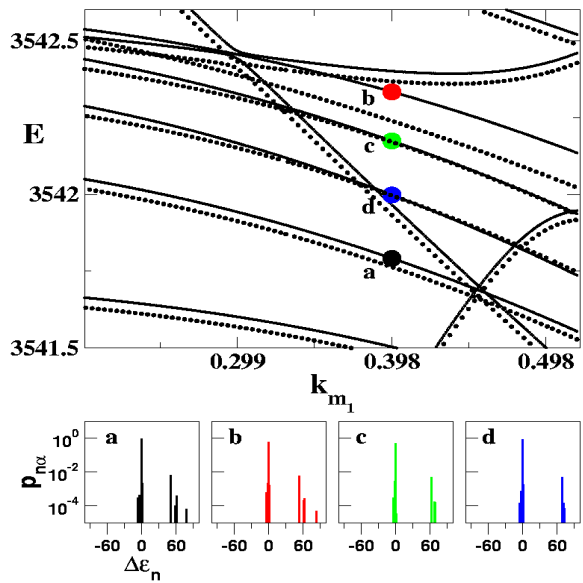}
\caption{Eigenvalues $E \equiv E_{\alpha}/\Delta E$ 
versus varying
coupling strength
$k_{{\bf m}_{1}} \equiv K_{{\bf m}_{1}}/\Delta E$.
$K_{{\bf m}_{2}} = K_{{\bf m}_{3}} \approx 0.2 \Delta E$.
The circles correspond
to the eigenstates having the largest contribution
from the specific zeroth-order states (Fig.~\ref{fig1}).
The dotted lines show a similar calculation using eq.~\ref{mqham}
indicating no qualitative change in the level motions.
The bottom panel
shows the overlap intensities
$p_{{\bf n}\alpha}=|\langle \alpha|{\bf n} \rangle|^{2}$
versus
$\Delta \epsilon_{\bf n} \equiv (E_{\bf n}^{0}-E_{\alpha})/\Delta E$.
Note the log-scale for the intensities and the
cluster of lines in all the plots around $\Delta \epsilon_{\bf n}=0$, and $60$.}\label{fig2}
\end{figure}

Specifically, we investigate a set of zeroth-order states around
$\bar{E} \approx 3542.5 \Delta E$ and $P=8$. 
This choice of $\bar{E}$ is 
motivated by the existence of a number of near-degenerate
states and qualitatively similar behavior is seen at different
values of $\bar{E}$ as well.
We select states that are not directly coupled 
by the perturbations in eq.~\ref{qham} but nevertheless
have IPR smaller than one. 
In Fig.~\ref{fig1} we show the survival probabilities for four such
zeroth-order states $|a\rangle = |0,11,1,4\rangle$,
$|b\rangle = |0,11,2,3\rangle$, $|c\rangle = |0,12,2,2\rangle$, and
$|d\rangle = |0,13,1,2\rangle$ with $\sigma_{\bf n}=0.97,0.51,0.40$, and
$0.74$ respectively. The crucial thing to note is
that $|b\rangle$,$|c\rangle$, and $|d\rangle$ mix amongst
themselves over long times. The corresponding
classical calculations, shown in Fig.~\ref{fig1} middle row,
indicate long time trapping. Thus the observed mixing between the
states is classically forbidden and corresponds to dynamical
tunneling. 
In Fig.~\ref{fig2} the variation of the
energy levels with the coupling parameter $k_{{\bf m}_{1}}$ is 
shown to indicate the lack of 
avoided crossings between the states
of interest. 
Fig.~\ref{fig2} also shows the
spectral intensities $p_{{\bf n}\alpha}$ and
in every case we see two clumps of lines - one at the origin and
another $\approx 60 \Delta E$ away. A purely
quantum explanation invokes the second clump of states, the virtual
or off-resonance states, 
which provide a "vibrational superexchange"
pathway for the mixing\cite{super}. 
The virtual states themselves have $\sigma_{\bf n} \approx 1$ and
hence do not mix significantly.
It is particularly striking to note that neither
$p_{{\bf n}\alpha}$ nor the energy level variations suggest
any differences between the zeroth-order states in
contrast to the observations in Fig.~\ref{fig1}. 
We now show that 
a relatively simple explanation can be
given in terms of resonance-assisted tunneling on the energy shell.

\begin{figure}
\includegraphics*[height=2.5in,width=3.0in]{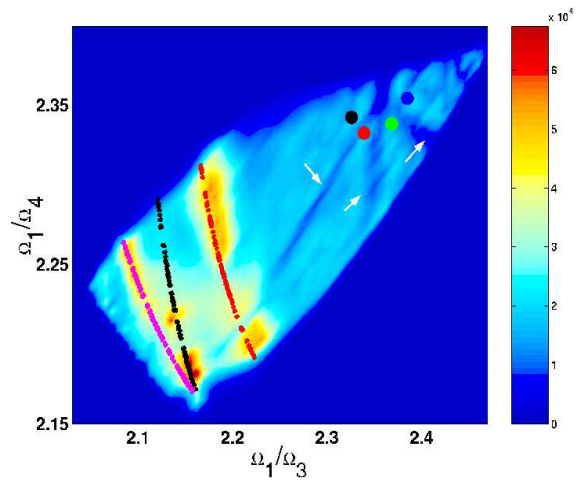}
\caption{The ``dynamical'' Arnol'd web at $E=\bar{E}$
generated by propagating $25000$ classical
trajectories for $\tau \approx 40$.
The primary resonances, ${\bf m}_{1}$ (red), ${\bf m}_{2}$ (black) and
${\bf m}_{3}$ (magenta),
as predicted
by ${\cal H}_{0}$ are superimposed for comparison.
${\bf m}_{2}$ and ${\bf m}_{3}$
intersect on the energy shell giving rise to the induced
resonance ${\bf m}_{i1} = (0,0,1,-1)$.
${\bf m}_{2}$ and ${\bf m}_{3}$ interact with
${\bf m}_{1}$ to give rise to the two other induced
resonances ${\bf m}_{i2}=(0,1,-1,0)$ and ${\bf m}_{i3}=(0,1,0,-1)$
respectively.
The nearly degenerate zeroth-order states (filled circles)
are located close to the
induced resonances (arrows) which lead to
state-mixing (cf. Fig.~\ref{fig1}) via
resonance-assisted tunneling.}
\label{fig3}
\end{figure}

In the resonance-assisted tunneling scenario the mixing between, for
example, 
$|b\rangle$ and $|c\rangle$ can be mediated by
a 1:1 resonance involving modes $j=2$ and $4$ {\it i.e.,} a
resonance vector ${\bf m}_{i3}\equiv(0,1,0,-1)$. 
The Hamiltonian in eq.~\ref{cham} does
not have ${\bf m}_{i3}$ explicitly but it can
be induced by ${\bf m}_{1}$ and
${\bf m}_{3}$.
Similarly, ${\bf m}_{i1}\equiv(0,0,1,-1)$ and
${\bf m}_{i2}\equiv(0,1,-1,0)$ can be induced by the resonances
in eq.~\ref{cham}. The resonances can be visualized by
constructing the Arnol'd web\cite{licht} at $E\approx \bar{E}$
and fixed $P$,
{\it i.e.,} the intersection of the various resonance planes 
${\bf m}_{r} \cdot \partial {\cal H}_{0}({\bf I})/\partial {\bf I} =0$ 
with 
the energy shell ${\cal H}_{0}({\bf I}) \approx \bar{E}$. For near-integrable
systems the energy shell, resonance zones, and
the location of the zeroth-order states can be projected onto
a $2$-dimensional space of two independent frequency ratios. 
The ``static'' Arnol'd web
based on ${\cal H}_{0}$ highlights the 
various possible resonances and their
topology on ${\cal H}_{0}({\bf I}) \approx \bar{E}$. 
However, from the tunneling
perspective it is crucial to determine 
dynamically relevant part of the static web at $E \approx \bar{E}$.  
This ``dynamical web'' is determined via a wavelet based
local frequency analysis\cite{arewig} 
of the Hamiltonian eq.~\ref{cham}. Briefly,
initial conditions $({\bf I},{\bm \theta})$ satisfying 
${\cal H}({\bf I},{\bm \theta}) \approx \bar{E}$ 
are generated and the trajectories
are followed in the frequency ratio space 
$(\Omega_{1}/\Omega_{3},\Omega_{1}/\Omega_{4})$.
The frequencies $\Omega_{k}(t)$ are 
computed along the trajectory $({\bf I},{\bm \theta})(t)$
by performing the
wavelet transform of $z_{k}(t)=\sqrt{2I_{k}(t)}\exp(i\theta_{k}(t))$.
The total number of
visits in a given region of the ratio space gives a density plot representing
the web which highlights dynamically significant regions
at a given energy. The resulting "dynamical web" is shown in
Fig.~\ref{fig3} along with the location of the relevant zeroth-order states.
Apart from highlighting the primary resonances ${\bf m}_{1},{\bf m}_{3}$ 
the figure also indicates the existence of the induced resonances
${\bf m}_{i1},{\bf m}_{i2}$, and ${\bf m}_{i3}$
separating the states. The states are
located close to the junction formed by the three induced
resonances and far away from the primary resonances. 
Hence Fig.~\ref{fig3} supports the notion that
$|b\rangle,|c\rangle$, and $|d\rangle$ 
are mixed due to dynamical tunneling mediated by
the induced resonances. 

\begin{figure}
\includegraphics*[height=2.0in,width=2.5in]{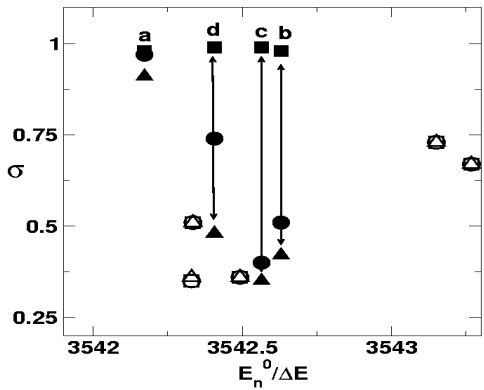}
\caption{IPR (eq.~\ref{ipr}) of the near-degenerate states
with respect to
$H$ (circles), $H'$ with the correct sign (squares), and
$H'$ with the opposite sign (triangles). 
$|\lambda_{{\bf m}_{i2}}| \approx |\lambda_{{\bf m}_{i3}}|
\approx 0.014 \Delta E$ in eq.~\ref{mqham}.
Note the importance of
the sign of the
induced resonances and the shutdown of tunneling for the states
of interest (filled symbols).
Other nearby states (open symbols) are least affected
by the counter-resonant terms.} 
\label{fig4}
\end{figure}

In order to conclusively establish the
role of the induced resonances it is necessary to extract their
strengths  
by perturbatively\cite{rat2,ozo,self} removing the primary
resonances in eq.~\ref{cham} to $O(\epsilon)$. 
As a result we obtain the
effective Hamiltonian containing the induced resonances
at $O(\epsilon^{2})$ which are approximated   
by effective pendulums.
For instance, one obtains the
effective pendulum Hamiltonian
\begin{equation}
{\cal H}^{(24)}_{eff} = \frac{(K_{24}-K_{24}^{r})^{2}}{2M_{24}} +
2 V_{24} \cos 2\phi_{24},
\end{equation}
appropriate for the induced resonance $\lambda_{{\bf m}_{i3}}$
with $K_{24} \sim 2 I_{4}$ and $2\phi_{24} \sim (\theta_{2}-\theta_{4})$.
The resonance center is denoted by $K_{24}^{r}$.
The coupling $V_{24}$ can be expressed in terms
of the conserved quantities $I_{3},P_{c}$, and $P_{24} \equiv I_{2}+I_{4}$
and the resulting tunneling time  
$\tau_{tun} \approx (2\pi \Delta E/V_{24})$ 
agrees well with Fig.~\ref{fig1}.
The strengths estimated via a pendulum approximation
can be translated back to effective quantum strengths $\lambda_{{\bf m}}$
and our analysis reveals that the induced resonances
are more than an order of magnitude smaller then that of the primary
resonances 
($\lambda_{{\bf m}_{i1}} \ll |\lambda_{{\bf m}_{i2}}| 
\approx |\lambda_{{\bf m}_{i3}}|
\approx 0.035 K_{{\bf m}_{1}}$). 
Note that ${\bf m}_{i2}$ and ${\bf m}_{i3}$
come with a negative sign as opposed to ${\bf m}_{i1}$. 
It is known\cite{rat2} that for significant mixing the states must lie
symmetrically with respect to the center of the mediating resonance zone.
Among the states considered, $|b\rangle$
and $|c\rangle$ satisfy the criterion very well and hence 
enhanced mixing between them is seen in Fig.~\ref{fig1}. The
state $|a\rangle$ is not symmetrically located with respect
to $|b\rangle$ and thus, combined with
the very small strength $\lambda_{{\bf m}_{i1}}$,
the induced resonance ${\bf m}_{i1}$ is
ineffective. 
Now consider
modifying eq.~\ref{qham} according to
\begin{equation}
H'=H+|\lambda_{{\bf m}_{i2}}|(a_{2}^{\dagger} a_{3} + {\rm h.c.})+
|\lambda_{{\bf m}_{i3}}|(a_{2}^{\dagger} a_{4} + {\rm h.c.}),
\label{mqham}
\end{equation}
where we have added terms to counter
the induced resonances. The reasoning is simple - if the
induced resonances are truly mediating the dynamical tunneling
then adding the counter-resonances should suppress the tunneling.
Moreover, since 
$\lambda_{{\bf m}_{i2}},\lambda_{{\bf m}_{i3}}\ll K_{{\bf m}_{1,2,3}}$
quantities like mean level spacing, eigenvalue variations (cf. Fig.~\ref{fig2})
and spectral intensities show little change as compared to the 
original system. Despite this, as shown in Fig.~\ref{fig1} bottom row,
the survival probabilities for
$|b\rangle,|c\rangle$, and
$|d\rangle$ indicate an almost complete shutdown of dynamical
tunneling. In Fig.~\ref{fig4} we show that the IPR also
indicate the shutdown of tunneling whereas IPR
of other nearby states 
do not change with the inclusion
of the counter terms. The result of
adding the counter terms with the same magnitude but opposite sign
shows (cf. Fig.~\ref{fig4})
enhanced mixing and this
establishes that the induced resonances 
${\bf m}_{i2}$ and ${\bf m}_{i3}$ are responsible for
the dynamical tunneling seen in Fig.~\ref{fig1}. 

In conclusion, this work shows that significant mixing between near-degenerate
states due to resonance-assisted tunneling can be expected in
very general situations. In addition, by suitable local modifications of
the phase space, complete control of the dynamical tunneling can
be attained. In light of a recent work the counter-resonances
can be thought of as weak control terms\cite{chand} and in nonautonomous
systems this suggests the possibility of control via additional weak
driving fields with particular attention to the
relative phases between the fields\cite{cont}.
The model system studied herein is certainly not
in the deep semiclassical limit, perhaps reason enough to argue against
competition from classical transport mechanisms,
and yet the importance of the nonlinear
resonances is clear. Further work
in the deep semiclassical regime, more closely approaching the
molecular systems, is in progress.

Part of this work was done at the Max-Planck-Institut f\"{u}r
Physik Komplexer Systeme, Dresden and I am grateful to Prof. J. M. Rost
for the hospitality and support. 
I thank P. Schlagheck for inspiring discussions
and A. Semparithi for
generating the data for Fig.~\ref{fig3}.


\begin{thebibliography}{}
\bibitem{davhel}{M. J. Davis and E. J. Heller, J. Chem. Phys. {\bf 75},
246 (1981); R. T. Lawton and M. S. Child, Mol. Phys. {\bf 37}, 1799 (1979).}
\bibitem{cats}{O. Bohigas, S. Tomsovic, and D. Ullmo, Phys. Rep. {\bf 223},
43 (1993); S. Creagh, in {\em Tunneling in Complex Systems}, edited by 
S. Tomsovic (World Scientific, Singapore, 1998), p.1.}
\bibitem{cat1}{W. A. Lin and L. E. Ballentine, 
Phys. Rev. Lett. {\bf 65}, 2927 (1990);
A. Shudo and K. S. Ikeda, Phys. Rev. Lett. {\bf 74}, 682 (1995);
V. A. Podolskiy and E. E. Narimanov, Phys. Rev. Lett. {\bf 91}, 263601 (2003);
S. Tomsovic and D. Ullmo, Phys. Rev. E {\bf 50}, 145 (1994);
J. Zakrzewski, D. Delande, and A. Buchleitner, Phys. Rev. E {\bf 57},
1458 (1998).}
\bibitem{frido}{E. Doron and S. D. Frischat, Phys. Rev. Lett. {\bf 75},
3661 (1995).}
\bibitem{rat1}{L. Bonci {\it et al}., 
Phys. Rev. E {\bf 58}, 5689 (1998).}
\bibitem{rat2}{ O. Brodier, P. Schlagheck, and 
D. Ullmo, Phys. Rev. Lett. {\bf 87}, 064101 (2001);
Ann. Phys. (N. Y.) {\bf 300}, 
88 (2002)}
\bibitem{ozo}{A. M. Ozorio de Almeida, J. Phys. Chem. {\bf 88}, 6139
(1984).}
\bibitem{ejhsar}{E. J. Heller and M. J. Davis, J. Phys. Chem. {\bf 85},
307 (1981); E. J. Heller, J. Phys. Chem. {\bf 99}, 2625 (1995).}
\bibitem{elt}{C. Eltschka and P. Schlagheck, Phys. Rev. Lett. {\bf 94},
014101 (2005).}
\bibitem{self}{S. Keshavamurthy, J. Chem. Phys. {\bf 119}, 161 (2003);
{\bf 122}, 114109 (2005) and references therein to the 
work on dynamical tunneling and energy flow in the molecular context.}
\bibitem{expt}{J. U. N\"{o}ckel and A. D. Stone, Nature {\bf
385}, 45 (1997); W. K. Hensinger {\it et al}.,
Nature {\bf 412}, 52 (2001); D. A.
Steck, W. H. Oskay, and M. G. Raizen, Science {\bf 293}, 274 (2001);
A. P. S. de Moura {\it et al.},
Phys. Rev. Lett. {\bf 88}, 236804 (2002); E. R. Th. Kerstel {\it et al.},
J. Phys. Chem. {\bf 95}, 8282 (1991).}
\bibitem{excep}{Effects of asymmetry are studied in, S. Tomsovic,
J. Phys. A: Math. Gen. {\bf 31}, 9469 (1998).}
\bibitem{licht}{A. J. Lichtenberg and M. A. Lieberman,
{\em Regular and Chaotic Dynamics}, Springer-Verlag, New York, 1992.}
\bibitem{quack}{$H_{0}$ corresponds to the chiral
molecule CDBrClF and the parameters are 
determined (table VIII, column 4) in, A. Beil {\it et al.},
J. Chem. Phys. {\bf 113}, 2701 (2000).}
\bibitem{simil1}{One can think of eq.~\ref{qham} as being in the
intrinsic resonance representation. See, M. Carioli, E. J. Heller, and
K. B. Moller, J. Chem. Phys. {\bf 106}, 8564 (1997); D. M. Leitner
and P. G. Wolynes, Phys. Rev. Lett. {\bf 76}, 216 (1996).}
\bibitem{super}{A. A. Stuchebrukhov and R. A. Marcus,
J. Chem. Phys. {\bf 98}, 8443 (1993); {\bf 98}, 6044 (1993).}
\bibitem{arewig}{L. V. Vela-Arevalo and S. Wiggins, 
Int. J. Bifur. Chaos {\bf 11}, 1359 (2001).}
\bibitem{chand}{C. Chandre {\it et al.}, Phys. Rev. Lett. {\bf 94},
074101 (2005).}
\bibitem{cont}{F. Grossmann {\it et al.,} Phys. Rev. Lett. {\bf 67}, 
516 (1991); D. Farrelly and J. A. Milligan, Phys. Rev. E. {\bf 47}, 
R2225 (1993);
M. Latka, P. Grigolini, and B. J. West, 
Phys. Rev. E {\bf 50}, R3299 (1994); V. Averbukh {\it et al.,} Z. Phys. D
{\bf 35}, 247 (1995).}
\end{thebibliography}
\end{document}